**Record Low Thermal Conductivity of Polycrystalline $MoS_2$ films: Tuning the Thermal Conductivity by Grain Orientation**

M. Sledzinska[1], R. Quey[2], B. Mortazavi[3], B. Graczykowski[4,5], M. Placidi[6], D. Saleta Reig[1,7], D. Navarro-Urrios[1], F. Alzina[1], L. Colombo[8], S. Roche[1,9] and C. M. Sotomayor Torres[1,9]

[1] Catalan Institute of Nanoscience and Nanotechnology (ICN2), CSIC and The Barcelona Institute of Science and Technology, Campus UAB, Bellaterra, E-08193 Barcelona, Spain
[2] Ecole des Mines de Saint-Étienne, CNRS UMR 5307, 158 cours Fauriel, F-42023 Saint-Étienne, Cedex 2, France
[3] Advanced Materials Multiscale Modeling, Institute of Structural Mechanics, Bauhaus-Universität Weimar, Marienstr. 15, D-99423 Weimar, Germany
[4] Max Planck Institute for Polymer Research, Ackermannweg 10, 55128, Mainz ,Germany
[5] NanoBioMedical Centre, Adam Mickiewicz University, ul. Umultowska 85, PL-61614 Poznan, Poland
[6] Catalonia Institute for Energy Research (IREC), Jardíns de les Dones de Negre 1, E-08930, Sant Adrià de Besòs, Spain
[7] Departament de Física, Universitat Autònoma de Barcelona, E-08193 Bellaterra, Spain
[8] Dipartimento di Fisica, Università di Cagliari, Cittadella Universitaria, I-09042 Monserrato (Ca), Italy
[9] ICREA—Institució Catalana de Recerca i Estudis Avancats, E-08010 Barcelona, Spain

**Abstract**

We report a record low thermal conductivity in polycrystalline $MoS_2$ obtained by varying grain sizes and orientations in ultrathin films. By optimizing the sulphurisation parameters of nanometre-thick Mo layer, we could grow $MoS_2$ films with tuneable morphologies. The thermal conductivity is extracted from a Raman laser power-dependent study on suspended samples. The lowest value of thermal conductivity of 0.27 $Wm^{-1}K^{-1}$, which reaches a similar value as teflon, is obtained in a polycrystalline sample formed by a combination of horizontally and vertically oriented grains, with respect to the bulk (001) monocrystal. Analysis by means of molecular dynamics and finite element method simulations confirm that such grain arrangement leads to lower grain boundary conductance. We discuss the possible use of these thermal insulating films in the context of electronics and thermoelectricity.

Recent advances in two-dimensional (2D) material science have shown the importance of phonon thermal transport in managing heat in future devices[1]. In particular, the understanding of how the thermal conductivity of polycrystalline 2D materials scales with grain sizes has become a key information for engineering efficient and scalable materials towards applications[2, 3]. At the same time research on thermoelectric effect in 2D materials has grown, as it was realized that low dimensional materials offer new routes to efficient thermoelectric generation because of the high power factor due to the unique density of states of confined charge carriers[4]. In fact, it was already shown that confinement of electrons or holes in structures such as quantum-well super-lattices or quantum-wires can largely enhance the thermoelectric conversion[4-6].

Various theoretical works have suggested efficient thermoelectric generation based on $MoS_2$ [7-10] which, together with progress in transition metal dichalcogenides (TMDs) doping, could lead to validation of these concepts in the near future[11, 12] . Crystalline $MoS_2$ films currently display thermal conductivity ranging from 34.5 $Wm^{-1}K^{-1}$ for a single layer[13] to 52 $Wm^{-1}K^{-1}$ for a few-layer[14]. Recently, the authors measured thermal conductivity below 1 $Wm^{-1}K^{-1}$ in polycrystalline thin films[15]. In common with other types of van der Waals materials, $MoS_2$ films also exhibit strong anisotropy between in-plane and cross-plane thermal conductivity[16]. However, to date the obtained values remain too high to envision $MoS_2$ as a thermal insulator or as an efficient part of thermoelectric generators. Here we show an efficient way of further reducing the thermal conductivity by modulating the grain orientation in ultrathin $MoS_2$ films.

Samples were prepared by means of sulphurisation of thin Mo films at high temperatures (see Experimental Methods) resulting in a set of samples with thickness of 3, 5, 8, 10 and 15 nm. The samples were transferred on holey $Si_3N_4$ substrates with 10 µm holes (Norcada Inc.) and onto transmission electron microscopy (TEM) grids using a polymer-free, wet transfer technique[15]. This particular transfer method was used to minimize possible polymer-related contamination. The thickness of the samples on the original $SiO_2$ substrate was measured by atomic force microscopy (see Supporting Information).

The TEM measurements revealed the change in the crystallographic structure of the samples with increasing thickness, as shown in Fig. 1. The thinnest (3nm) sample, which corresponds to approximately 4 layers of $MoS_2$, shows horizontally oriented grains (HG), aligned with a predominant (100) orientation (Fig. 1(a) and 1(f)). For the thickest sample (15nm) we observe almost fully vertically oriented grains (VG) with predominant (002) orientation, as shown in the Fig 1(e) and 1(g). The other samples (Fig. 1 (b)-(d)) exhibit a combination of both horizontal and vertical grains and the presence of the latter increases with the sample thickness.

The synthesis of horizontally and vertically oriented TMD microcrystals by RF sputtering[17, 18] and thin metal layer sulphurisation[19, 20] was studied in the late 80s and early 90s for applications in dry lubrication or catalysis. Recently, Kong et al. demonstrated the growth of fully vertically oriented $MoS_2$ and $MoSe_2$ thin films[21]. The explanation offered for this arrangement was that, in these films mass transport along the layers through van der Waals gaps is much faster than across the layers. Therefore, the layers tend to be oriented perpendicular to the substrate, ensuring exposed van der Waals gaps for faster reaction[21]. Our experiments further validate this model. As the thickness of the sample increases, sulphur penetration into the Mo layer is favoured by layers oriented perpendicular to the film.

Raman scattering of bulk transition metal dichalcogenides has been studied since the 70s, resulting in the identification of two principal first-order modes: $A_{1g}$ (410 $cm^{-1}$) and $E_{2g}$ (382 $cm^{-1}$)[22, 23] . In Fig. 2(a) we compare the Raman spectra taken on the samples on their original $SiO_2$ substrates. The Raman spectra of $MoS_2$ nanosheets were obtained at the same incident laser power (0.3 mW) and integrated for the same amount of time (180 s). As expected, the $A_{1g}$ peak is shifted towards lower frequencies as the nanosheet thickness increases. Furthermore, the Raman peaks of the nanosheet are broader and less intense with increasing proportion of vertically oriented grains. The asymmetric broadening of the line shape, which is already evident in the sample with predominant HGs, is attributed to the small grain size[24, 25]. The decrease of the intensity of both Raman peaks is related to its dependence on the orientation of the grains, either horizontally or vertically, displaying no distinct texture and probing a large number of grains within the laser spot (of about 1 μm)[26].

Both $MoS_2$ Raman peaks are temperature dependent, shifting to lower frequencies as the temperature increases. The Raman shift as a function of temperature was measured in our previous work, with the temperature dependence of the Raman coefficients of $\chi_{E2g}$ = -1.40·10$^{-2}$ cm$^{-1}$K$^{-1}$ for the $E_{2g}$ mode and $\chi_{A1g}$ = -1.22·10$^{-2}$ cm$^{-1}$K$^{-1}$ for the $A_{1g}$ mode[15]. Fig 2(b) shows schematically the Raman thermometry setup used for thermal conductivity measurements[13, 14]. The measurements were performed at room temperature and in vacuum. The laser power-dependent Raman spectra were obtained on suspended, circular samples with diameter of 10μm (inset to Fig. 2(c)). With increasing laser power the Raman-active modes soften due to the local temperature increase of the $MoS_2$. In what follows we concentrate on the $A_{1g}$ mode, as it is more intense than the $E_{2g}$ mode. Using the temperature Raman coefficient we can then calculate the temperature of the hotspot $T_p$ as a function of the absorbed power $P_{abs}$. As an example, the data for the 3 nm thick sample is shown in Fig. 2(c), together with the linear fit.

To extract the thermal conductivity from the experimental data we used a FEM model (developed in COMSOL Multiphysics) describing the temperature distribution as a function of the absorbed power, illustrated schematically in Fig. 2(d). The solution of the problem is given by simulating Fourier's equation:

$$P_{abs}/A = -\kappa \nabla T \qquad (1)$$

Where $A$ is the cross-sectional area of the heat flux, $\kappa$ is the thermal conductivity, and $T$ is the temperature. The laser heating/probing spot on the sample was treated as a uniform heat source along the direction normal to the sample surface and Gaussian in the radial direction:

$$P(r) = P_0 \exp(-2r^2/w_0^2) \qquad (2)$$

where $r$ is the distance from the membrane centre, $P_0$ is the Gaussian amplitude defined by the total absorbed power $P_{abs}$ and Gaussian beam waist size $w_0$, as follows $P_0 = 2P_{abs}/(\pi w_0^2)$. Due to the symmetry of the membrane and its isotropic in-plane thermal conductivity the model was simplified to a 2D stationary heat flow study. The temperature $T_p$ probed at the sample was approximated by the formula:

$$T_p = \frac{\int_0^{r_m} T(r)P(r)r\mathrm{d}r}{\int_0^{r_m} P(r)r\mathrm{d}r} \qquad (3)$$

Where $T(r)$ is the temperature distribution obtained from the FEM simulation and $r_m$ is the membrane radius. Here for the temperature-independent $\kappa$ and in the absence of radiative and convective losses $T_p$ is a linear function of $P_{abs}$. Therefore, the thermal conductivity of the measured sample is determined by an iterative procedure where $\kappa$ is swept as a model parameter until $T_p(P_{abs})$ fits the experimental data and the calculated one matches with highest possible numerical accuracy. The calculated temperature profile of the 3 nm sample is shown in Fig. 2 (d) and (e).

The measured thermal conductivity of the five samples is shown in Fig. 3(a) as a function of the proportion of the vertical grains (black dots). For the 3 nm thick sample, composed only of horizontally-oriented grains, the thermal conductivity was extracted to be of 2.0 ± 0.2 Wm$^{-1}$K$^{-1}$, which is comparable to our previous results in a similar type of $MoS_2$ sample[15]. However, as the proportion of vertical grains in the sample increases we observe the decrease of $\kappa$ down to 0.27 ± 0.15 $Wm^{-1}K^{-1}$ for the 10 nm thick sample, which contains similar proportions of vertical and horizontal grains. A further increase of $\kappa$ to 0.55 ± 0.15 $Wm^{-1}K^{-1}$ is measured for the 15 nm thick sample, in which grains are mostly vertically-oriented.

To understand the observed thermal transport features, we developed a multiscale approach, in which atomistic simulations provide $\kappa$ values for single $MoS_2$ grain with varying size, while 2D finite element simulations are used to calculate the overall $\kappa$ of the polycrystalline samples.

The first task was accomplished by non-equilibrium molecular dynamics simulations (NEMD), implemented by the LAMPPS package[27] and based on the force field developed by Kandemir et al.[28] for $MoS_2$. The equations of motion were integrated by the Verlet velocity method with a time-step as short as 0.2 fs, while the temperature control was operated by means of Nosé–Hoover thermostat.

Due to the anisotropy of the grains in the $MoS_2$ thin films (see Fig. 1) we set a distinction between the thermal conductivity parallel to the basal plane, $\kappa_{||}$, and the thermal conductivity normal to the basal plane, $\kappa_{\perp}$, with respect to the bulk (001) monocrystal. The normal-to-plane conductivity is known to be of the order of $\kappa_{\perp} \sim 2$ $Wm^{-1}K^{-1}$ [29], and we considered this value independent on the grain size.

In Fig. 3(b) red symbols represent the NEMD results. As expected smaller grains have smaller $\kappa_{||}$ values, but the trend is clearly approaching an upper limit, which we identify as the basal plane thermal conductivity $\kappa_{\infty}$ of an infinite and defect-free $MoS_2$ sample. In order to quantify this trend, we can fit the set of NEMD data by

$$\frac{1}{k_1} = \frac{1}{k_\infty}\left(1 + \frac{\Lambda}{d}\right) \tag{4}$$

where $\Lambda$ is the effective phonon mean free path[30, 31]. The fitting procedure yields $\kappa_{\infty} = 53.6$ $Wm^{-1}K^{-1}$ and $\Lambda = 80.8$ nm. Although the result for $\kappa_{\infty}$ is underestimated with respect to previous predictions based on first-principles solution of the Boltzmann transport equation (BTE)[32], the agreement between NEMD and BTE data for $\kappa_{||}$ (L), where L represents a range of grain lengths containing the actual grain size distribution of the measured samples is excellent for $L \leq 50$ nm (Fig. 3 (b)). This provides support to the reliability of the following investigation in polycrystalline systems.
By inserting the actual experimental value into eq.(4) for the typical grain length L ~ 2 -12 nm of the measured samples, we obtain a value of $\kappa_{||} \sim$ 1-7 $Wm^{-1}K^{-1}$. Interestingly, this value is of the same order as $\kappa_{\perp} \sim 2$ $Wm^{-1}K^{-1}$ which implies that for the grain sizes here considered only a moderate anisotropy is expected.

We now turn to the polycrystalline $MoS_2$ case, the first task being the construction of a structural model. To this aim, we switched to the continuum, finite element description and generated microstructures from the experimental grain size and orientation distributions using the Neper software package[33], as done in a previous study [15], but this time with two grain populations (HGs and VGs). In the left panel of Fig. 3(c) we show a typical polycrystalline sample the thermal transport properties of which have been further investigated by FEM analysis. While less fundamental and accurate than atomistic simulation, this approach allows the incorporation of the structural complexity of the samples studied. Unlike atomistic simulations, FEM simulations of heat transport in polycrystalline $MoS_2$ make it possible to establish a robust correlation between the underlying nanostructure and the resulting physical properties. For each experimental sample, a microstructure was generated from the grain size distributions of Fig. 1, and each grain was assigned an anisotropic thermal conductivity, with $\kappa_{||}(L)$ taken from Fig. 3(b).

When considering a polycrystalline $MoS_2$, we must duly include the role grain boundaries (GBs) play in heat transport. In particular, we need to assign a thermal conductance to each GB and we must consider all possible relative alignments, namely, two adjacent grains could be both horizontally- or both vertically-oriented, or there can be one horizontal and one vertical grain. Hereafter, we will label the three possibilities as hh, vv, or hv, respectively. This results in three GB thermal conductances, namely $h_{hh}$, $h_{hv}$, and $h_{vv}$ to be determined with the FEM simulations.

Since the GB conductance is unknown, we explored three possible combinations, by setting: (i) $h_{hh} = h_{hv} = h_{vv}$ ; (ii) $h_{hh} \neq h_{hv} = h_{vv}$ ; (iii) $h_{hh} \neq h_{hv} \neq h_{vv}$. In case (i), the FEM simulation provides only a limited decrease of the thermal conductivity as the proportion of vertically-oriented grains increases, contrary to experimental findings.

If all GBs have the same conductance, the modest reduction can only be attributed to the change of the grain conductivities. In case (ii), simulations provide a monotonic decrease of thermal conductivity as the proportion of vertically-oriented grains increases, once again in contrast to experimental data. Finally, in case (iii), FE simulations reproduce well experimental results. In particular, the thermal conductivity increase with higher proportions of vertically-oriented grains is well reproduced. The results are shown in Fig. 3(a), together with the actual fitted values of $h_{hh}$, $h_{hv}$, and $h_{vv}$.

Interestingly, in the most realistic situation corresponding to case (iii), $h_{hv}$ is two orders of magnitude smaller than $h_{vv}$ and $h_{hh}$, both resulting to be of the same order. This result indicates that the density of interfaces between horizontally and vertically-oriented grains is the controlling factor for the thermal conductivity.

One of the possible explanations is that the grain boundaries between the HGs and VGs are not perfectly defined due to the growth process and they may contain some amorphous matter, which decreases strongly the GB conductance as a nm-thin interfacial layer. Figure 3(c) shows that this high contrast in thermal conductance results in a highly heterogeneous thermal diffusion in such microstructures.

To summarize, we have reported an ultralow thermal conductivity of 0.27 $Wm^{-1}K^{-1}$ in thin films of polycrystalline $MoS_2$ with a disordered distribution of horizontally and vertically aligned grains. This opens promising prospects for thermoelectric energy conversion. Indeed, one observes that even for a thermal conductivity up to two to three orders of magnitude larger in crystalline $MoS_2$ films, theoretical predictions put the value of ZT as high as 0.11 at 500 K [9]. Moreover, measurements in a 16 nm thick film of 2D $SnS_2$ crystal reported a room temperature thermal conductivity of 3.45 $Wm^{-1}K^{-1}$ with a ZT=0.13, i.e., two orders of magnitude larger than the bulk three-dimensional $SnS_2$[34]. Our ultrathin films of $MoS_2$, down to 4 monolayers, exhibit a one order of magnitude lower thermal conductivity compared with 2D $SnS_2$ suggesting that, given the similarities in their electronic properties, further improvement of ZT using ultrathin $MoS_2$ should be possible.

## Experimental Methods

*$MoS_2$ synthesis*

The synthesis of the $MoS_2$ nanosheets was achieved via a thin-film conversion technique. A thin Mo layer with varying thickness was deposited by DC-magnetron sputtering on 5x5 $cm^2$ glass substrates and reactively annealed in a graphite box in a sulphur containing atmosphere at 600ºC for 30 minutes.

*$MoS_2$ wet transfer*

The free-standing structures were fabricated using an etching- and polymer-free, surface tension-assisted wet transfer method[15].

The CVD-grown material was directly submerged in DI water to detach it from its substrate. The different surface energies drove water molecules to penetrate underneath the $MoS_2$ film, which then floated on the water surface. Holey $Si_3N_4$ membranes (Norcada Inc.) were used to scoop the floating $MoS_2$ from the water surface. The samples were dried on a hotplate at 100◦C

*$MoS_2$ characterisation*

Atomic force microscopy topography images were obtained using a Nanoscope IV controller and a Dimension 3100 head (Veeco).

High-resolution imaging of structure and morphology of the samples was obtained using FEI Tecnai F20 in TEM and STEM modes. The grain distribution analysis was performed using Digital Micrograph and ImageJ software.

*Raman thermometry*

The Raman scattering measurements were performed in a Linkam temperature controlled vacuum stage (THMS350V) under vacuum at room temperature. A Horiba T64000 Raman spectrometer and a 532 nm laser (Cobolt Samba) were used to obtain the spectra.

The laser beam was focused on the sample with the microscope objective (50x and NA = 0.55) acting as a Gaussian heat source with a waist size of about 1 µm. The absorbed power $P_{abs}$ is measured for each sample as the difference between incident and transmitted plus reflected laser power. The powers are measured with a calibrated system based on cube non-polarizing beam splitters, i.e., no assumptions are made of the sample optical absorption. The measurements were performed in vacuum ($5x10^{-3}$ mPa) in order to reduce convective loses.

*Modelling*

In the NEMD for transport in the basal plane, a suitable thermal bias was applied across single-grain $MoS_2$ samples by thermostating its opposite terminations at temperatures $T_{hot}$=310 K and $T_{cold}$=290 K, respectively, and by imposing periodic boundary conditions along the direction transverse to heat transport. The systems were aged until a steady-state transport regime was established, after about 100 ps of heat exchanging process. Eventually, the room temperature basal plane thermal conductivity $\kappa_{||}$ was evaluated as the ratio between the calculated heat flux and the corresponding temperature gradient established across the sample. This quantity was evaluated by only considering the central part of the simulated sample, sufficiently away from its thermostated terminal ends, where the temperature profile was found to be linear[35]. The calculation was repeated for 10 samples, differing in their total length L, which, was varied in the range 5 nm $\leq$ L $\leq$ 100 nm to cover the actual experimental range. This procedure has provided reliable results in a previous investigation of nanocrystalline $MoS_2$ systems[35].

For the finite element simulations microstructures were generated respecting the experimental volume fractions and size distributions of horizontally and vertically oriented grains, using the Neper software package [33]. For each sample, 1000 grains were used and attributed crystal orientations according to their family: horizontally oriented grains were assigned random orientations with crystal direction 3 perpendicular to the sample plane while vertically-oriented grains were assigned random orientations with crystal direction 3 contained in the sample plane. The details of the structures used in the simulations are shown in the supplementary information.

## Acknowledgements

The authors acknowledge the financial support from the FP7 project QUANTIHEAT (No. 604668); the Spanish MICINN project PHENTOM (FIS2015-70862-P); and the program Severo Ochoa (Grant SEV-2013-0295). B.G. acknowledges the support from the Homing Programme granted by the Foundation for Polish Science and the support from the Alexander von Humboldt foundation. S. R. was supported by the European Union Horizon 2020 research and innovation programme under grant agreement No. 696656 (Graphene Flagship), the Spanish Ministry of Economy and Competitiveness and the European Regional Development Fund [Project No. FIS2015-67767-P (MINECO/FEDER)] and by the Universities and Research Secretariat of the Ministry of Business and Knowledge of the Generalitat de Catalunya. D.N-U acknowledges the support of a Ramón y Cajal postdoctoral fellowship (RYC-2014-15392).

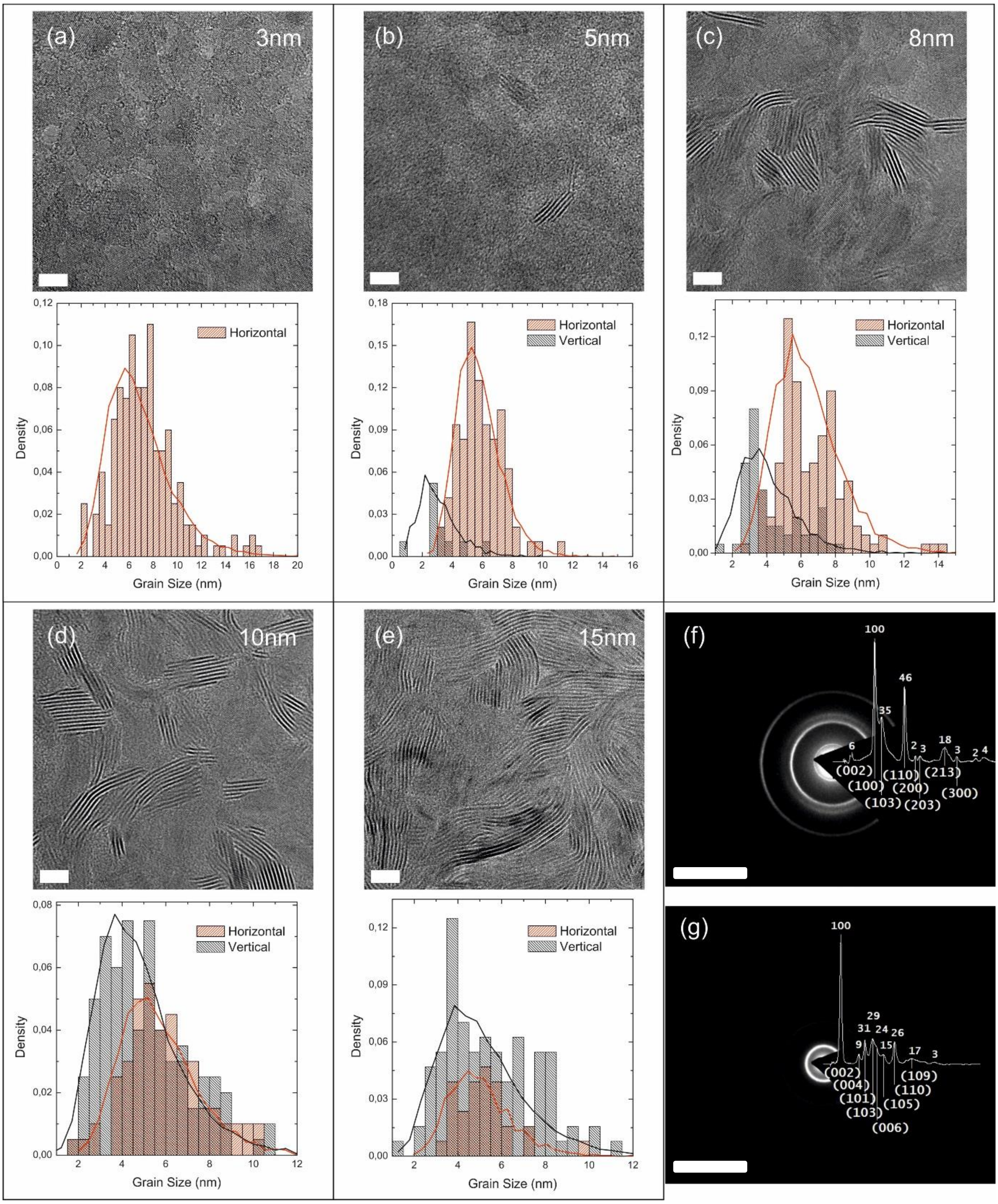

Figure 1. TEM images and grain distribution of samples of (a) 3nm, (b) 5nm, (c) 8nm, (d) 10nm and (e) 15nm thickness. Scale bars correspond to 5 nm. Solid lines in the grain size distribution correspond to the grain distribution used in the simulations. Electron diffraction pattern of sample thickness of (f) 3 nm and (g) 15 nm. Scale bar corresponds to 5 $nm^{-1}$.

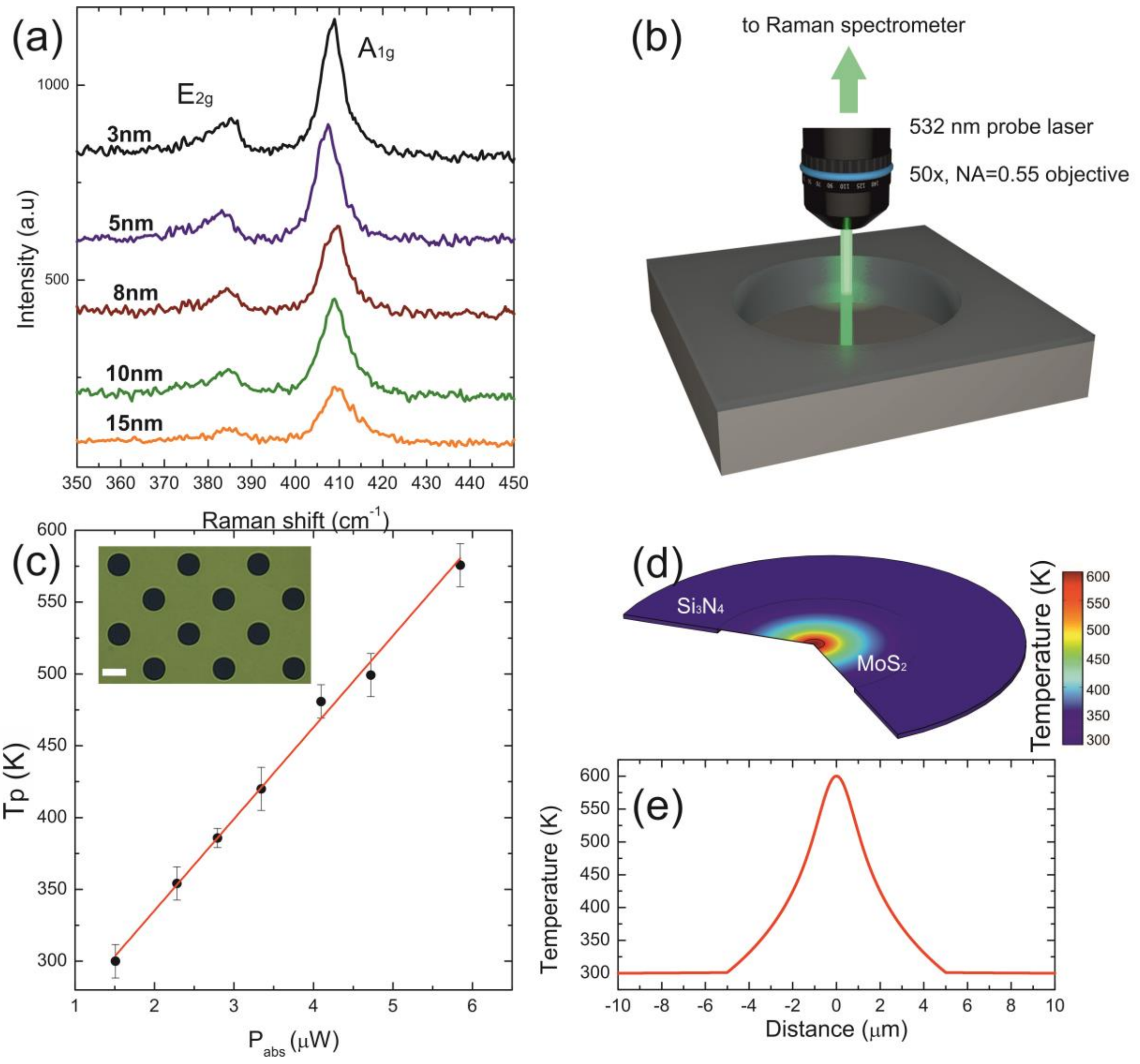

Figure 2. Raman thermometry of $MoS_2$ ultrathin films: (a) Characteristic Raman peaks $A_{1g}$ (410 $cm^{-1}$) and $E_{2g}$ (382 $cm^{-1}$) of the $MoS_2$ nanosheets on the original $SiO_2$ substrate. (b) Illustration of the Raman laser thermometry set up. (c) Increase of sample temperature as a function of absorbed power $P_{abs}$ and the linear fit to the experimental data. (d) COMSOL simulation of the temperature distribution in 3nm –thick sample for a given absorbed power of 5 μW. (e) Temperature profile as a function of distance from the hot spot for a given absorbed power of 5 μW, simulated using COMSOL.

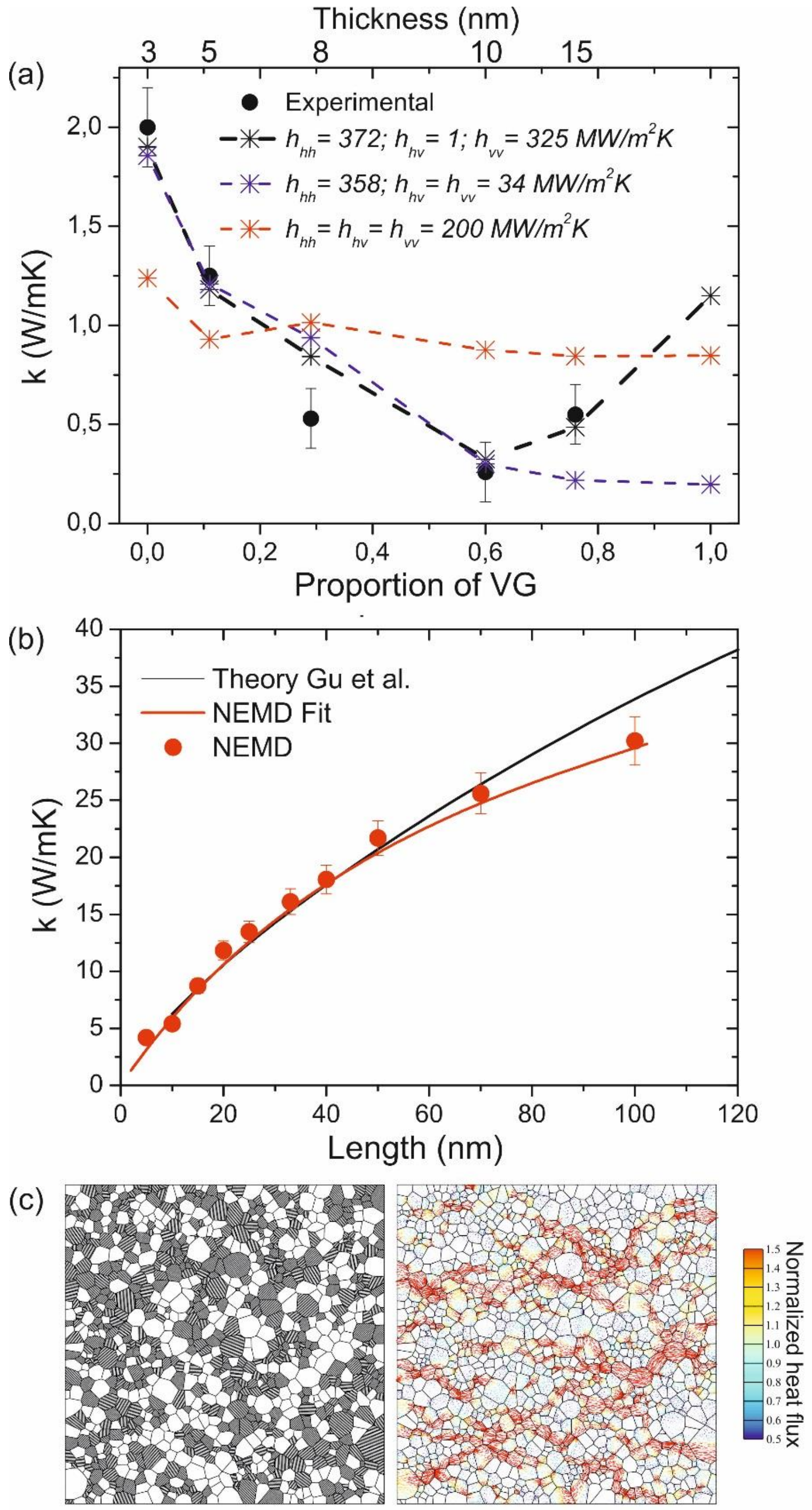

Figure 3. Effect of grains in the thermal conductivity of $MoS_2$. (a) Thermal conductivity as a function of vertical grains proportion. Experimental results are shown in black dots and simulations results for three different assumptions of grain boundary conductance in coloured stars. Dashed lines are guide for the eye. The thickness of each sample is indicated on the upper axis. (b) Size-dependence of the thermal conductivity of a single-layer $MoS_2$ at 300 K calculated by NEMD. The red and black solid lines are the fitting of the NEMD results and theoretical results of Gu et al.[32], respectively. (c) Schematics of grain distribution, where grey and white correspond to vertically and horizontally

oriented grains, respectively (left) and thermal diffusion in a 10 nm thick sample simulated by finite elements methods assuming $h_{hh}$ = 372 MW/m$^2$K, $h_{hv}$ = 1 MW/m$^2$K and $h_{vv}$ = 325 MW/m$^2$K (right).